\documentclass[twocolumn,aps,prl,floatfix,superscriptaddress]{revtex4-2}

\usepackage{hyperref}
\usepackage[utf8]{inputenc}

\RequirePackage{lineno}

\usepackage{graphicx}
\usepackage{dcolumn}
\usepackage{amsmath}
\usepackage{supertabular}
\usepackage{multirow}
\usepackage{color}
\usepackage{cancel}
\usepackage{ulem}
\usepackage{array}
\usepackage{tabularx}

\usepackage{amsmath}
\usepackage{amssymb}
\usepackage{xcolor}
\definecolor{myOrange}{rgb}{1,0.5,0.}
\definecolor{myGreen}{rgb}{0.0,0.6,0.1}
 
\newcommand{\removetext}[1]{}

\newcommand{\aaa}{\ensuremath{A\text{--}A}}
\newcommand{\AuAu}{\ensuremath{\mathrm{Au}\text{--}\mathrm{Au}}}
\newcommand{\PbPb}{\ensuremath{\mathrm{Pb}\text{--}\mathrm{Pb}}}
\newcommand{\pp}{\ensuremath{pp}}

\newcommand{\sqrtsNN}{\ensuremath{\sqrt{s_\mathrm{NN}}}}
\newcommand{\sqrts}{\ensuremath{\sqrt{s}}}

\newcommand{\pizero}{\ensuremath{\pi^0}}

\newcommand{\gammadir}{\ensuremath{\gamma_\mathrm{dir}}}
\newcommand{\gammarich}{\ensuremath{\gamma_\mathrm{rich}}}

\newcommand{\pT}{\ensuremath{p_\mathrm{T}}}

\newcommand{\pTtrack}{\ensuremath{p_\mathrm{T,track}}}

\newcommand{\pTjet}{\ensuremath{p_\mathrm{T,jet}}}

\newcommand{\pTraw}{\ensuremath{p_\mathrm{T,jet}^\mathrm{raw}}}

\newcommand{\pTreco}{\ensuremath{p_\mathrm{T,jet}^\mathrm{reco,ch}}}

\newcommand{\Ajet}{\ensuremath{A_\mathrm{jet}}}

\newcommand{\rhoA}{\ensuremath{\rho\Ajet}}

\newcommand{\ET}{\ensuremath{E_\mathrm{T}}}
\newcommand{\ETtrig}{\ensuremath{E_\mathrm{T}^\mathrm{trig}}}

\newcommand{\pTtrig}{\ensuremath{p_\mathrm{T}^\mathrm{trig}}}
\newcommand{\pTjetch}{\ensuremath{p_\mathrm{ T,jet}^\mathrm{ch}}}

\newcommand{\gev}{\ensuremath{\mathrm{GeV/}c}}

\newcommand{\zvtx}{\ensuremath{z_\mathrm{vtx}}}

\newcommand{\antikT}{\ensuremath{\mathrm{anti-}k_\mathrm{T}}}

\newcommand{\rr}{\ensuremath{R}}

\newcommand{\dphi}{\ensuremath{\Delta\phi}}
\newcommand{\deta}{\ensuremath{\Delta\eta}}

\newcommand{\etajet}{\ensuremath{\eta_\mathrm{jet}}}
\newcommand{\etatrack}{\ensuremath{\eta_\mathrm{track}}}

\newcommand{\Ntrig}{\ensuremath{N_\mathrm{trig}}}

\newcommand{\AAtoTrig}{\ensuremath{AA}\rightarrow\mathrm{trig}}
\newcommand{\AAtoTrigjet}{\ensuremath{AA\rightarrow\mathrm{trig+jet}}}
\newcommand{\dNjetdpTddphi}{\ensuremath{\frac{{\rm d}^{2}N_{\mathrm{jet}}}{\mathrm{d}\pTjetch\mathrm{d}\dphi}}}

\newcommand{\IAA}{\ensuremath{{I}_\mathrm{AA}}}
\newcommand{\IAADphi}{\ensuremath{{I}_\mathrm{AA}(\Delta\phi)}}
\newcommand{\IAApT}{\ensuremath{{I}_\mathrm{AA} ({\pTjetch})} }

\newcommand{\Qsq}{\ensuremath{{Q}^{2}}}

\newcommand{\qhat}{\ensuremath{\hat{q}}}
\newcommand{\qhatL}{\ensuremath{\langle\qhat{L}\rangle}}

\newcommand{\YpTdphi}{\ensuremath{Y(\pTjetch,\dphi)}}

\newcommand{\YpT}{\ensuremath{Y(\pTjetch)}}
\newcommand{\YpTreco}{\ensuremath{Y(\pTreco)}}
\newcommand{\Ydphi}{\ensuremath{Y(\dphi)}}

\newcommand{\Ytilde}{\ensuremath{\widetilde{Y}}}
\newcommand{\Ytildedphi}{\ensuremath{\widetilde{Y}_{\dphi}}}
\newcommand{\YtildepT}{\ensuremath{\widetilde{Y}_{\pT}}}

\newcommand{\YSE}{\ensuremath{Y_\mathrm{SE}}}
\newcommand{\YME}{\ensuremath{Y_\mathrm{ME}}}
\newcommand{\YMEnorm}{\ensuremath{Y_\mathrm{MEnorm}}}

\begin{document}


\title{Measurement of medium--induced acoplanarity  in central \AuAu\ and \pp\ collisions at $\sqrtsNN=200$ GeV using direct--photon+jet and \pizero+jet correlations}

\bigskip

\affiliation{Academia Sinica, Nankang, 115, Taipei}
\affiliation{Abilene Christian University, Abilene, Texas   79699}
\affiliation{AGH University of Krakow, FPACS, Cracow 30-059, Poland}
\affiliation{Argonne National Laboratory, Argonne, Illinois 60439}
\affiliation{American University in Cairo, New Cairo 11835, Egypt}
\affiliation{Ball State University, Muncie, Indiana, 47306}
\affiliation{Brookhaven National Laboratory, Upton, New York 11973}
\affiliation{University of Calabria \& INFN-Cosenza, Rende 87036, Italy}
\affiliation{University of California, Berkeley, California 94720}
\affiliation{University of California, Davis, California 95616}
\affiliation{University of California, Los Angeles, California 90095}
\affiliation{University of California, Riverside, California 92521}
\affiliation{Central China Normal University, Wuhan, Hubei 430079 }
\affiliation{University of Illinois at Chicago, Chicago, Illinois 60607}
\affiliation{Chongqing University, Chongqing, 401331}
\affiliation{Creighton University, Omaha, Nebraska 68178}
\affiliation{Czech Technical University in Prague, FNSPE, Prague 115 19, Czech Republic}
\affiliation{Technische Universit\"at Darmstadt, Darmstadt 64289, Germany}
\affiliation{National Institute of Technology Durgapur, Durgapur - 713209, India}
\affiliation{ELTE E\"otv\"os Lor\'and University, Budapest, Hungary H-1117}
\affiliation{Frankfurt Institute for Advanced Studies FIAS, Frankfurt 60438, Germany}
\affiliation{Fudan University, Shanghai, 200433 }
\affiliation{Guangxi Normal University, Guilin, 541004}
\affiliation{University of Heidelberg, Heidelberg 69120, Germany }
\affiliation{University of Houston, Houston, Texas 77204}
\affiliation{Huzhou University, Huzhou, Zhejiang  313000}
\affiliation{Indian Institute of Science Education and Research (IISER), Berhampur 760010 , India}
\affiliation{Indian Institute of Science Education and Research (IISER) Tirupati, Tirupati 517507, India}
\affiliation{Indian Institute Technology, Patna, Bihar 801106, India}
\affiliation{Indiana University, Bloomington, Indiana 47408}
\affiliation{Institute of Modern Physics, Chinese Academy of Sciences, Lanzhou, Gansu 730000 }
\affiliation{University of Jammu, Jammu 180001, India}
\affiliation{Kent State University, Kent, Ohio 44242}
\affiliation{University of Kentucky, Lexington, Kentucky 40506-0055}
\affiliation{Lanzhou University, Lanzhou, 730000}
\affiliation{Lawrence Berkeley National Laboratory, Berkeley, California 94720}
\affiliation{Lehigh University, Bethlehem, Pennsylvania 18015}
\affiliation{Lovely Professional University, Jalandhar - Delhi G.T. Road, Pagwara, Panjab, 144411, India}
\affiliation{Max-Planck-Institut f\"ur Physik, Munich 80805, Germany}
\affiliation{Michigan State University, East Lansing, Michigan 48824}
\affiliation{National Institute of Science Education and Research, HBNI, Jatni 752050, India}
\affiliation{National Cheng Kung University, Tainan 70101 }
\affiliation{Nuclear Physics Institute of the CAS, Rez 250 68, Czech Republic}
\affiliation{The Ohio State University, Columbus, Ohio 43210}
\affiliation{Panjab University, Chandigarh 160014, India}
\affiliation{Purdue University, West Lafayette, Indiana 47907}
\affiliation{Rice University, Houston, Texas 77251}
\affiliation{Rutgers University, Piscataway, New Jersey 08854}
\affiliation{University of Science and Technology of China, Hefei, Anhui 230026}
\affiliation{South China Normal University, Guangzhou, Guangdong 510631}
\affiliation{Sejong University, Seoul, 05006, Korea, Republic Of}
\affiliation{Shandong University, Qingdao, Shandong 266237}
\affiliation{Shanghai Institute of Applied Physics, Chinese Academy of Sciences, Shanghai 201800}
\affiliation{Southern Connecticut State University, New Haven, Connecticut 06515}
\affiliation{State University of New York, Stony Brook, New York 11794}
\affiliation{Instituto de Alta Investigaci\'on, Universidad de Tarapac\'a, Arica 1000000, Chile}
\affiliation{Temple University, Philadelphia, Pennsylvania 19122}
\affiliation{Texas A\&M University, College Station, Texas 77843}
\affiliation{Texas Southern University, Houston, Texas, 77004}
\affiliation{University of Texas, Austin, Texas 78712}
\affiliation{Tsinghua University, Beijing 100084}
\affiliation{University of Tsukuba, Tsukuba, Ibaraki 305-8571, Japan}
\affiliation{University of Chinese Academy of Sciences, Beijing, 101408}
\affiliation{United States Naval Academy, Annapolis, Maryland 21402}
\affiliation{Valparaiso University, Valparaiso, Indiana 46383}
\affiliation{Variable Energy Cyclotron Centre, Kolkata 700064, India}
\affiliation{Warsaw University of Technology, Warsaw 00-661, Poland}
\affiliation{Wayne State University, Detroit, Michigan 48201}
\affiliation{Wuhan University of Science and Technology, Wuhan, Hubei 430065}
\affiliation{Yale University, New Haven, Connecticut 06520}

\author{B.~E.~Aboona}\affiliation{Texas A\&M University, College Station, Texas 77843}
\author{J.~Adam}\affiliation{Czech Technical University in Prague, FNSPE, Prague 115 19, Czech Republic}
\author{L.~Adamczyk}\affiliation{AGH University of Krakow, FPACS, Cracow 30-059, Poland}
\author{I.~Aggarwal}\affiliation{Panjab University, Chandigarh 160014, India}
\author{M.~M.~Aggarwal}\affiliation{Panjab University, Chandigarh 160014, India}
\author{Z.~Ahammed}\affiliation{Variable Energy Cyclotron Centre, Kolkata 700064, India}
\author{A.~K.~Alshammri}\affiliation{Kent State University, Kent, Ohio 44242}
\author{D. M. Anderson}\affiliation{Texas A\&M University, College Station, Texas 77843}
\author{E.~C.~Aschenauer}\affiliation{Brookhaven National Laboratory, Upton, New York 11973}
\author{S.~Aslam}\affiliation{Fudan University, Shanghai, 200433 }
% ...

\collaboration{STAR Collaboration}\noaffiliation

\noaffiliation



\begin{abstract}
The STAR Collaboration reports measurements of acoplanarity using semi--inclusive distributions of charged--particle jets recoiling from direct photon and \pizero\ triggers, in central \AuAu\ and \pp\ collisions at $\sqrtsNN=200$ GeV. Significant medium--induced acoplanarity broadening is observed for large but not small recoil jet resolution parameter, corresponding to recoil jet yield enhancement up to a factor of $\approx20$ for trigger--recoil azimuthal separation far from $\pi$. This phenomenology is indicative of the response of the Quark--Gluon Plasma to excitation, but not the scattering of jets off of its quasiparticles. The measurements are not well--described by current theoretical models which incorporate jet quenching.
\end{abstract}

\maketitle
 

\section{Introduction} Matter under extreme conditions of temperature and density forms a state of matter called Quark--Gluon Plasma (QGP) which consists of deconfined quarks and gluons~\cite{Collins:1974ky,Shuryak:1977ut,Busza:2018rrf,Harris:2024aov}. A QGP filled the early universe a few microseconds after the Big Bang, and is generated in collisions of heavy nuclei at the Relativistic Heavy--Ion Collider (RHIC) and the Large Hadron Collider (LHC). Experimental measurements at these facilities, and their comparison to theoretical model calculations, show that the QGP exhibits emergent collective behavior, flowing with the lowest possible specific shear viscosity~\cite{Heinz:2013th}.

Lattice Quantum Chromodynamics (QCD) calculations of high--temperature matter at zero net--baryon density indicate that the effective number of degrees of freedom in the QGP is about 15\% less than the Stefan--Boltzmann limit for a non-interacting quark--gluon gas, even at temperatures several times the pseudo--critical temperature $T_\mathrm{c}\approx155$ MeV~\cite{Borsanyi:2013bia,Bhattacharya:2014ara,HotQCD:2014kol,Ding:2015ona,Bazavov:2017dus}. This indicates that QGP quasiparticles in this temperature range are complex multi--particle states of quarks and gluons, which may drive its collective dynamics ~\cite{Busza:2018rrf}. However, the microscopic structure of the QGP remains largely unexplored experimentally.

In high--energy hadronic collisions, quarks and gluons (partons) in the projectiles can experience hard (high momentum--transfer \Qsq) scattering. The scattered parton is initially virtual, decaying in a parton shower which hadronizes as an observable spray of hadrons (a ``jet'')~\cite{Abelev:2006uq,Adamczyk:2016okk,Abelev:2013fn,Aad:2014vwa,Khachatryan:2016mlc}. In high-energy nuclear collisions, hard scatterings occur before the formation of the QGP, and scattered partons subsequently interact with it (``jet quenching'')~\cite{Majumder:2010qh,Cunqueiro:2021wls,Apolinario:2022vzg}. Jet quenching modifies observed jet production rates and substructure, providing unique probes of the QGP~\cite{Cunqueiro:2021wls,Apolinario:2022vzg}.

The secondary scattering of hard partons in the QGP has been proposed as a probe of QGP quasiparticles~\cite{Appel:1985dq,Blaizot:1986ma,DEramo:2010wup,DEramo:2012uzl,Chen:2016vem,DEramo:2018eoy}, in analogy to Rutherford scattering as a probe of the atomic nucleus~\cite{Rutherford:1911zz}. Observation of secondary, in--medium partonic scattering may be observable using coincidence channels, in which a trigger particle associated with an initial hard scattering specifies a direction, and the azimuthal difference \dphi\ between a recoil jet and the trigger (acoplanarity with respect to the plane defined by the beam axis and trigger) is measured. Acoplanarity distributions have been measured at the LHC for \pp\ and central \PbPb\ collisions~\cite{ALICE:2015mdb,ALICE:2023qve,ALICE:2023jye,CMS:2012ulu,CMS:2012ytf,CMS:2017eqd,CMS:2017ehl}. Measurements by the CMS Collaboration utilize a direct--photon (\gammadir) trigger with transverse energy $\ET>40$ GeV and recoil jets with transverse momentum $\pTjet>30$ \gev, with jet resolution parameter $\rr=0.3$; no medium--induced modification is observed within uncertainties~\cite{CMS:2012ytf,CMS:2017ehl}. Likewise, no acoplanarity broadening is observed by CMS in high-\pT\ dijet~\cite{CMS:2011iwn} and Z+jet~\cite{CMS:2017eqd} correlations.

The ALICE Collaboration reports semi-inclusive distributions of charged--particle recoil jets with $\pTjetch>10$ \gev\ recoiling from hadron triggers with $\pTtrig>20$ \gev\ (h+jet); significant medium--induced broadening of the acoplanarity distribution is observed in the range $10<\pTjetch<20$ \gev\ for $\rr=0.4$ and 0.5, but not for $\rr=0.2$ or at higher \pTjetch~\cite{ALICE:2023qve,ALICE:2023jye}. This marked dependence of the broadening on \rr\ and \pTjetch\ for such low-\pT\ jets suggests that it arises from response of the QGP medium to excitation by a jet (``wake''), rather than single hard  Rutherford--like scattering~\cite{ALICE:2023qve,ALICE:2023jye}. CMS also reports a medium-induced statistical excess in soft--particle yield at large angles relative to the axis of sub-leading jets in unbalanced di-jet pairs, though without their incorporation in jet reconstruction or measurement of acoplanarity~\cite{CMS:2016cvr}. At RHIC, the STAR collaboration has also reported an acoplanarity measurement based on semi--inclusive h+jet correlations in central and peripheral \AuAu\ collisions at $\sqrtsNN=200$ GeV, with no significant medium--induced broadening observed for $\rr=0.3$~\cite{STAR:2017hhs}. A search for in--medium partonic scattering has also been carried out using jet substructure~\cite{ALICE:2024fip}.

In this article, the STAR experiment reports the first measurement of jet acoplanarity of charged--particle jets recoiling from direct photon (\gammadir) and \pizero\ triggers in \pp\ and central \AuAu\ collisions at $\sqrtsNN=200$ GeV. The trigger particles have $11<\ETtrig<15$ GeV, with the semi--inclusive distribution of recoil jets reported in $10<\pTjetch<20$ \gev\ for $\rr=0.2$ and 0.5. Uncorrelated jet background yield is corrected using event mixing. This article extends the analysis reported in Refs.~\cite{STAR:2023ksv,STAR:2023pal}, to measure acoplanarity. 

The measurement of acoplanarity distributions with \gammadir\ and \pizero\ triggers in the same analysis provides systematic variation in the recoil--jet color charge and pathlength distributions~\cite{STAR:2023ksv,STAR:2023pal}, elucidating their influence on jet quenching effects. This measurement complements that reported by ALICE in Ref.~\cite{ALICE:2023qve}, exploring acoplanarity broadening using the same observable but with a collision system at markedly different \sqrtsNN, thereby probing sensitivity to variation in the QGP temperature and expansion dynamics~\cite{Shen:2013vja,JETSCAPE:2020mzn}. Theoretical model calculations incorporating jet quenching are also compared to the measurements.


\section{Detector, dataset, and analysis} 
Data for \pp\ and \AuAu\ collisions at $\sqrtsNN=200$ GeV were recorded during the 2009 and 2014 RHIC runs, respectively. The detector, datasets, triggering, offline event selection, and track reconstruction are described in Ref.~\cite{STAR:2023ksv}. Online event selection is based on high--energy single showers measured in the Barrel Electromagnetic Calorimeter (BEMC)~\cite{Beddo:2002zx}. Centrality of \AuAu\ collisions is determined offline using the uncorrected charged--particle multiplicity within pseudo--rapidity $|\eta|<0.5$; the 15\% highest--multiplicity (``central'') \AuAu\ collisions are selected for analysis. Events are further selected in both the \pp\ and central \AuAu\ datasets by requiring the presence of a \gammadir\ or \pizero\ candidate with $11<\ETtrig<15$ GeV. The integrated luminosity for the analysis is 23 pb$^{-1}$ and 3.9 nb$^{-1}$ for \pp\ and \AuAu\ collisions, respectively. 

High-\pT\ photon production in RHIC collisions arises from several sources~\cite{Owens:1986mp,David:2019wpt}: direct ($2\rightarrow2$) production (Compton, annihilation), fragmentation, and hadronic decays. Discrimination of $\gamma$ and $\pi^0$--induced showers utilizes EM shower shape measured in the BEMC and its Shower Maximum detector (BSMD)~\cite{STAR:2016jdz,STAR:2023ksv}. The purity of the resulting \pizero--tagged population is estimated from simulation to be greater than 95\%~\cite{STAR:2023pal}, while the photon--tagged population contains an admixture of \pizero\ and is labeled ``$\gamma$--rich'' (\gammarich). The \gammadir\ fraction of the \gammarich\ population, which is determined using the measured rate of nearby correlated charged hadrons, depends on collision system and \ETtrig\ and is in the range $\approx 40-80\%$~\cite{STAR:2023pal}. Correction based on this rate accounts for the hadronic decay component, and much but not all of the fragmentation photon contribution~\cite{STAR:2016jdz,STAR:2023pal}. This population is labeled \gammadir.

Jet reconstruction likewise follows Ref.~\cite{STAR:2023ksv}. Jets are reconstructed from charged--particle tracks in $|\etatrack|<1$ and $\pTtrack>0.2$ \gev\ using the \antikT\ algorithm with $\rr=0.2$ and 0.5, with E--scheme recombination and active ghost area 0.01~\cite{Cacciari:2011ma}. Jets whose centroid has $|\etajet|<1-\rr$ are accepted for analysis, and measured distributions are normalized to unit $\etajet$. The raw jet transverse momentum, \pTraw, is adjusted according to $\pTreco=\pTraw-\rhoA$, where $\rho$ is the event--wise estimated background \pT-density and \Ajet\ is the jet area ~\cite{Cacciari:2007fd,STAR:2023ksv}. This event--wise approximate correction is refined by the deconvolution of detector effects (``unfolding'').

The number of jet candidates as a function of \pTjet\ and \dphi\ is normalized by \Ntrig, the number of triggers,

\begin{align}
\YpTdphi &\equiv
\frac{1}{\Ntrig}\cdot\dNjetdpTddphi\Bigg\vert_{\pTtrig}
\label{eq:hJetDefinition} \\
&= \left(\frac{1}{\sigma^{\AAtoTrig}} \cdot
\frac{{\rm d}^{2}\sigma^{\AAtoTrigjet}}{\mathrm{d}\pTjetch\mathrm{d}\dphi}\right)
\Bigg\vert_{\pTtrig}.
\label{eq:hJetXsection}
\end{align}

\noindent
The distribution \YpTdphi\ is normalized per unit $\deta$, not shown. Single--differential projections are denoted \YpT\ and \Ydphi. Since the trigger distribution is inclusive, the resulting distribution in the absence of uncorrelated background is equal to the semi--inclusive ratio of hard cross sections (Eq.~\ref{eq:hJetXsection}), where AA denotes either \pp\ or \AuAu.

The measured recoil--jet yield in central \AuAu\ collisions has multiple contributions: correlated recoil jets from the same hard (high--\Qsq) scattering process which generates the trigger, corresponding to \YpTdphi; jets arising from other partonic scattering processes, which are uncorrelated with the trigger (multiple partonic interactions, or MPI); and combinatorial jets arising from the random combination of tracks generated by soft (low--\Qsq) processes. The jet yield in central \AuAu\ collisions at $\sqrtsNN=200$ GeV due to MPI is negligible~\cite{STAR:2017hhs}. However, the combinatorial yield in such collisions can be significant relative to the correlated signal, especially for large \rr\ at low \pTjetch. 

Correction of the raw recoil-jet yield to measure the \YpTdphi\ distribution is first carried out on 1-D \YpTreco\ raw distributions  binned in raw \dphi, following the procedures described in Ref.~\cite{STAR:2017hhs}. The corrected 1-D distributions as a function of \pTjetch\ are then combined to form a corrected 2-D distribution in (\pTjetch,\dphi) using weights that account for \dphi-smearing due to residual background. The correction steps are outlined below, with detail provided for the \dphi--smearing correction.


\begin{figure*}[htb!]
\centering
\includegraphics[width=0.5\textwidth]{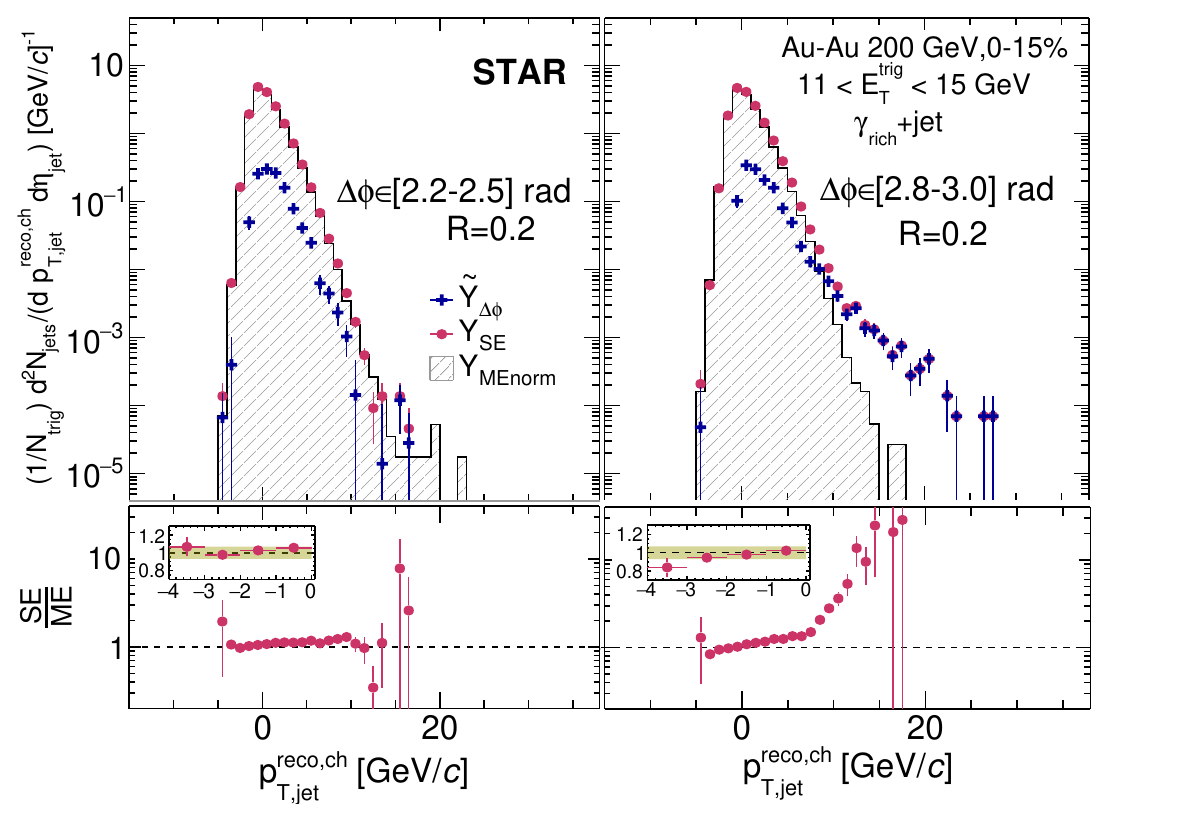}
\includegraphics[width=0.5\textwidth]{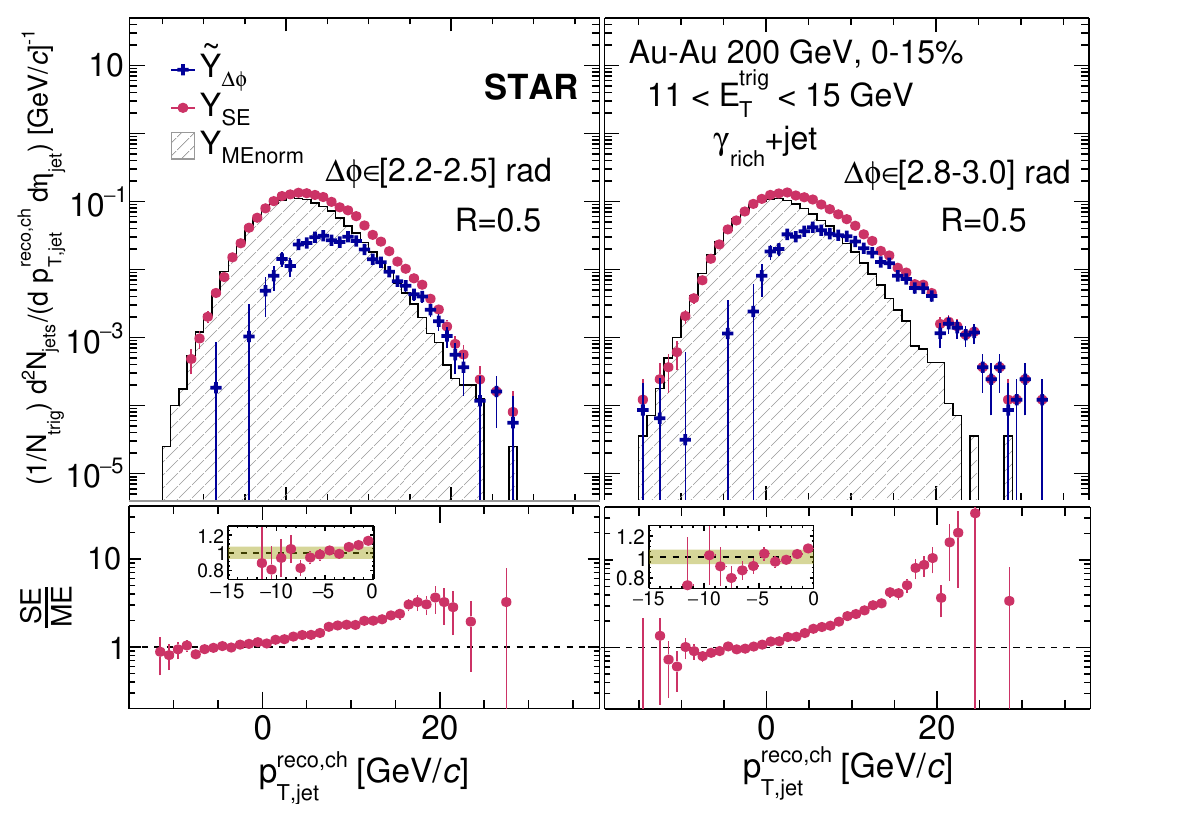}
\caption{\YSE, \YMEnorm\ and \Ytildedphi\ distributions (Eq.~\ref{eq:Ytilde}) as a function of \pTreco\ from \gammarich\ triggers with $11<\ETtrig<15$ GeV in central \AuAu\ collisions, for $\rr=0.2$ (top panels) and 0.5 (bottom panels). \Ytilde(\pTreco) datapoints with a negative central value not shown. Sub-panels for each \rr: $2.2<\dphi<2.5$ (left), $2.8<\dphi<3.0$ (right). Upper panels show SE and MEnorm distributions; lower panels show the ratio SE/MEnorm. Insets show the ratio in the normalization region.} 
\label{fig:SEME}
\end{figure*}

Correction for the combinatorial jet yield in central \AuAu\ collisions is carried out using mixed events (ME)~\cite{STAR:2017hhs,STAR:2023ksv}, which are constructed by combining single tracks from multiple real events (``same events,'' or SE) in each of 540 distinct bins in multiplicity, \zvtx\ (primary vertex beamline position), event plane orientation, and run--averaged luminosity~\cite{STAR:2023ksv}. Figure~\ref{fig:SEME} shows examples of the SE and ME (normalized, see below) distributions, and their ratio, for central \AuAu\ collisions with \gammarich\ triggers and recoil jets with $\rr=0.2$ and 0.5, in the ranges $2.2<\dphi<2.5$ and  $2.8<\dphi<3.0$. A negative value of \pTreco\ arises if \pTraw\ is less than \rhoA; for large negative values of \pTreco\ this occurs predominantly due to uncorrelated background, resulting in the same \pTreco--dependent shape of the \YSE\ and \YME\ distributions in that region~\cite{ALICE:2015mdb,STAR:2017hhs,ALICE:2023qve,ALICE:2023jye,STAR:2023ksv,STAR:2023pal}. 

The ME distribution is normalized to the SE distribution in the negative \pTreco\ region (MEnorm), which corrects the effective acceptance difference of the SE and ME populations due to displacement of uncorrelated jet candidates at low and negative \pTreco\ by hard, correlated jet candidates~\cite{ALICE:2015mdb,STAR:2017hhs}. The SE/ME yield ratio (lower panel insets) is independent of \pTreco\ within a few percent over a negative--\pTreco\ range in which the yield itself varies by several orders of magnitude; for $\rr=0.5$, the range in \pTreco\ over which the ratio is flat  within statistical uncertainty is limited. The extracted normalization factors have values between 0.9 and unity~\cite{STAR:2023ksv,STAR:2023pal}. The systematic dependence of the normalization factor on the upper bound of the normalization region is negligible as shown in the appendix Fig~\ref{fig:dphiCorrection}.

The distribution of recoil--jet yield correlated with the trigger corresponds to the difference distribution~\cite{ALICE:2015mdb,STAR:2017hhs}

\begin{equation}
\Ytildedphi(\pTreco)=\YSE(\pTreco)-\YMEnorm(\pTreco).
\label{eq:Ytilde}
\end{equation}

\noindent
The symbol \Ytildedphi\ denotes the distribution as a function of \pT\ in bins of \dphi, while \YtildepT\ (used below) denotes the distribution as a function of \dphi\ in bins of \pT. This data--driven, statistical  correction for uncorrelated yield enables recoil--jet measurements in central \aaa\ collisions over broad phase space, including low \pTjet\ and large \rr~\cite{STAR:2023ksv,STAR:2023pal,ALICE:2023qve,ALICE:2023jye}. No ME--based correction is applied for \pp\ collisions due to small background yield, {\it i.e.} \Ytilde(\pTreco)=\YSE(\pTreco).

Figure~\ref{fig:SEME} shows \Ytildedphi(\pTreco) distributions, which vary smoothly even for small signal/background yield, $\Ytilde\ll\YSE$. Negative values from the subtraction are not displayed due to the logarithmic vertical scale, but all such points have central values consistent with zero within statistical uncertainty. These features indicate that the ME distribution reproduces accurately the uncorrelated jet distribution in SE population, which can therefore be corrected with high precision~\cite{STAR:2017hhs}.

The $\Ytildedphi(\pTreco)$ distributions are then corrected via unfolding for instrumental effects in both \pp\ and central \AuAu\ collisions, and for residual uncorrelated background fluctuations in central \AuAu\ collisions~\cite{STAR:2023ksv,STAR:2017hhs}. Finally, the unfolded distributions are corrected for jet--finding efficiency~\cite{STAR:2023ksv}. The dominant systematic uncertainty is due to unfolding~\cite{STAR:2023ksv}. 

Instrumental effects generate negligible \dphi--smearing relative to the analysis binning. The only significant \dphi--smearing is due to spatial variation of uncorrelated background in central \AuAu\ collisions, which can modify the jet centroid direction. Correction for \dphi--smearing is implemented bin--wise by a weight matrix $w(\dphi_\mathrm{true},\dphi_\mathrm{meas})$ that scales the measured $\Ytildedphi(\pTjetch)$ distributions. The weights are determined by embedding  detector--level PYTHIA-generated events for \pp\ collisions into real central \AuAu\ events, with systematic uncertainty determined by varying the jet fragmentation model to mimic jet quenching effects. The systematic uncertainty of this correction is negligible as discussed in the appendix.

The \gammadir-triggered recoil--jet distributions for each \dphi\ bin are then determined from the corrected \gammarich--triggered and \pizero-triggered distributions~\cite{STAR:2023ksv}. 


\section{Results} 
\begin{figure}[htb!]
\includegraphics[width=0.5\textwidth]{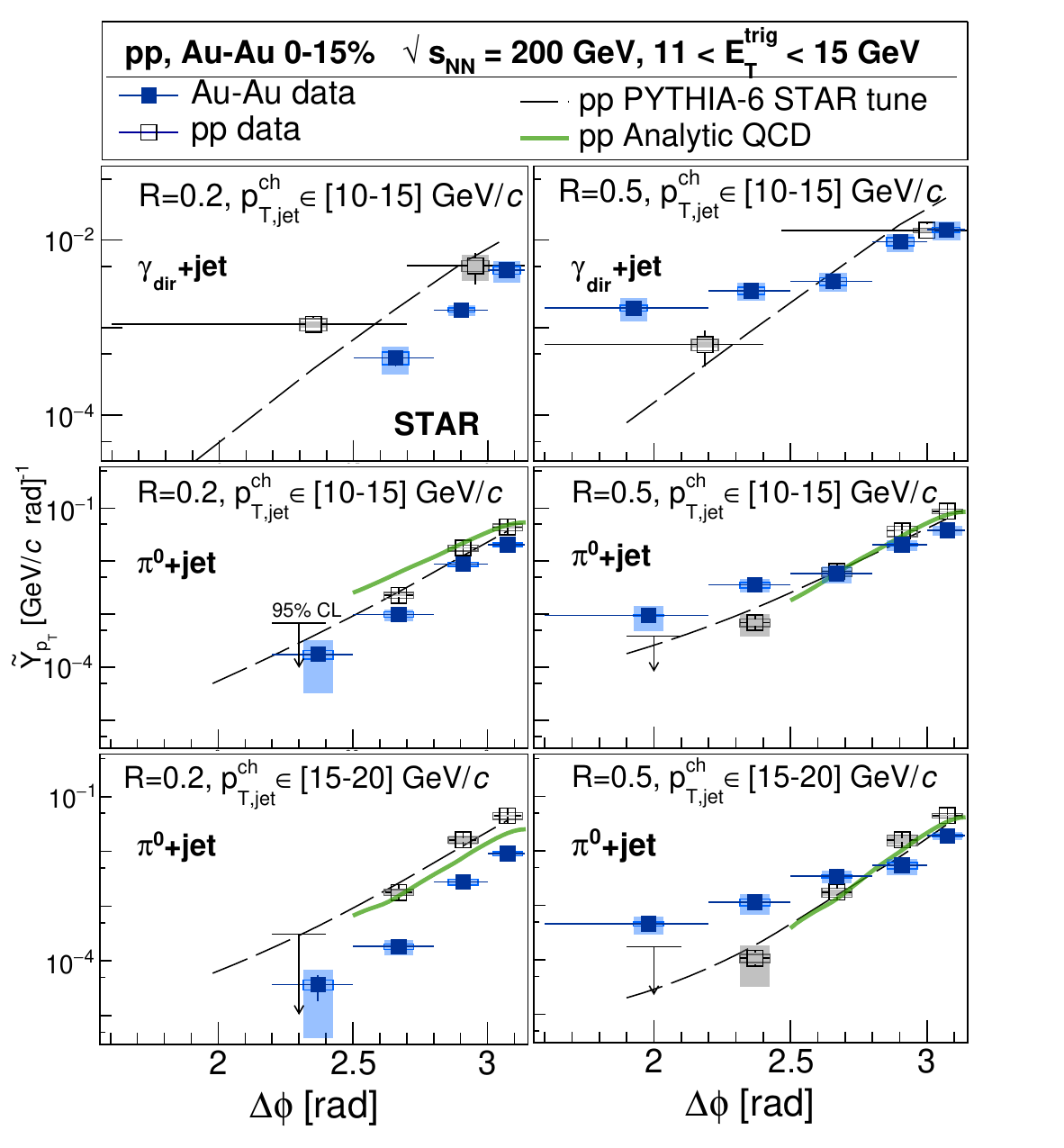}
\caption{Corrected \YtildepT\ distributions as a function of \dphi\ for \gammadir\ and \pizero\ triggers with $11<\ETtrig<15$ GeV, in \pp\ and central \AuAu\ collisions at $\sqrtsNN=200$ GeV for $\rr=0.2$ (left) and $\rr=0.5$ (right). Upper: $\gammadir$ trigger, $10<\pTjetch<15$ \gev; middle: $\pizero$ trigger, $10<\pTjetch<15$ \gev; lower: $\pizero$ trigger, $15<\pTjetch<20$ \gev. Data are plotted at the spectrum-weighted bin coordinate except for low--statistics points, which are shown as 95\% CL upper limits. Theoretical calculations for \pp\ collisions are described in the text.}
\label{fig:DelphiDist}
 \end{figure}

Figure~\ref{fig:DelphiDist} shows corrected \YtildepT(\dphi) distributions for \gammadir\ and \pizero\ triggers in \pp\ and central \AuAu\ collisions, for $\rr=0.2$ and 0.5. The distributions fall steeply away from $\dphi=\pi$, with greater yield for $\rr=0.5$ than for $\rr=0.2$. Figure~\ref{fig:SEME}, lower panels, shows greater yield for $\rr=0.5$ than for $\rr=0.2$ for $\pTreco>0$ at large angles relative to $\dphi=\pi$; this effect is therefore not generated predominantly by corrections. A similar effect is reported in Ref.~\cite{ALICE:2023qve}.

Figure~\ref{fig:DelphiDist} shows a calculation for \pp\ collisions at $\sqrts=200$ GeV using PYTHIA--6 STAR tune~\cite{Adam:2019aml}, with the \ETtrig\ distribution smeared to account for the BEMC detector response~\cite{STAR:2023ksv}. This calculation describes the measurements well for both \gammadir\ and \pizero\ triggers. The figure also shows an analytic QCD calculation at Next--to--Leading--Log (NLL) accuracy with Sudakov resummation~\cite{Sun:2014gfa,Chen:2016vem} for \pizero\ triggers in \pp\ collisions, in the range $2.5<\dphi<\pi$ rad. This calculation, which is not smeared by the \ETtrig\ resolution~\cite{STAR:2023ksv}, reproduces the \pp\ data well for $\rr=0.5$, but not for $\rr=0.2$. 

\begin{figure}[htb!]
\includegraphics[width=0.5\textwidth]{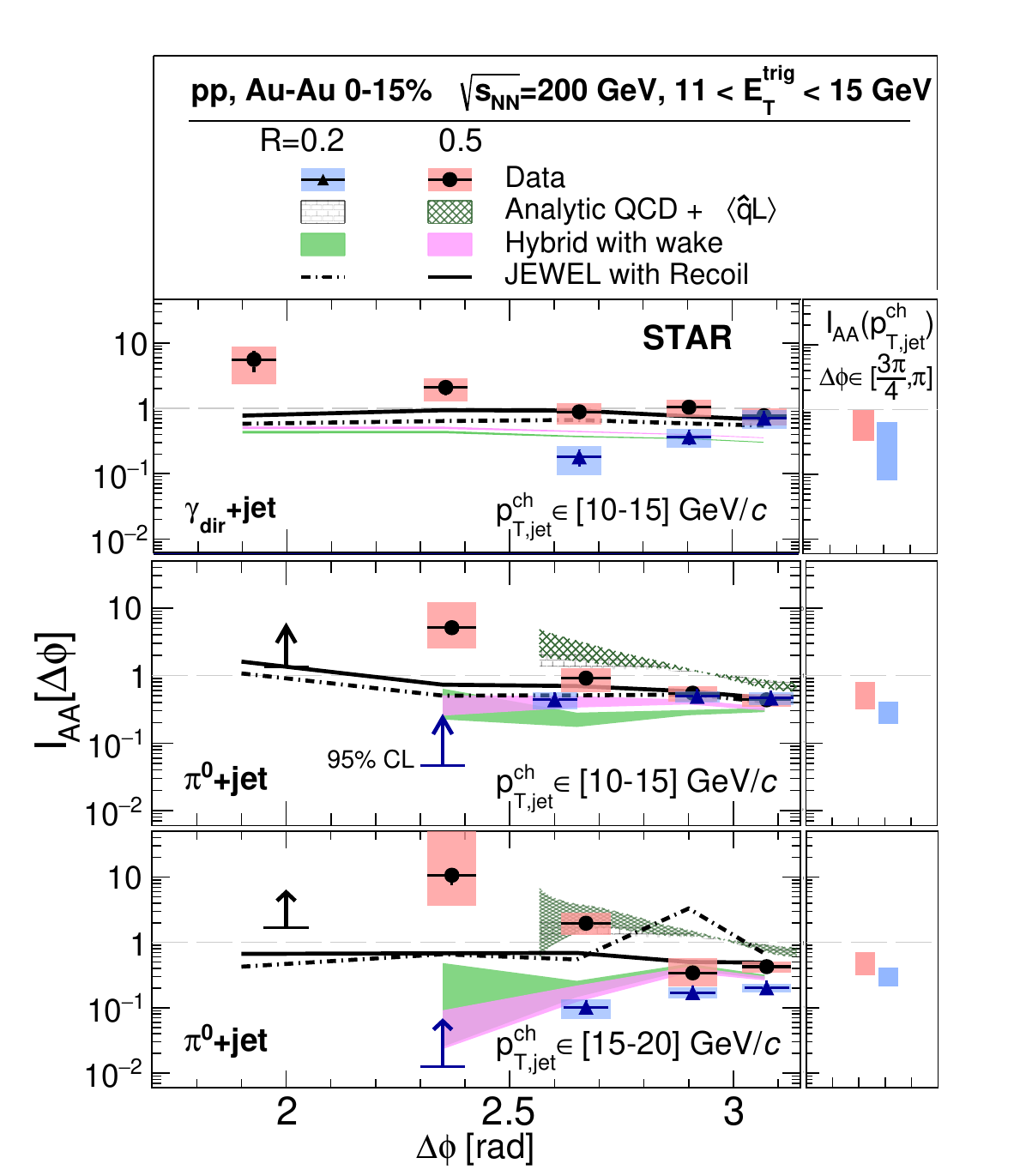}
\caption{Left panels: \IAADphi\ distributions for $11<\ETtrig<15$ GeV with recoil jet $\rr=0.2$ and 0.5. Top: \gammadir\ trigger, $10<\pTjetch<15$ \gev; middle: \pizero\  trigger, $10<\pTjetch<15$ \gev; bottom: \pizero\ trigger, $15<\pTjetch<20$ \gev. Arrows indicate 95\% CL. Theoretical calculations are discussed in the text. Right panels: integral of \IAADphi\ over $3\pi/4<\dphi<\pi$~\cite{STAR:2023ksv}.}
\label{fig:IAA}
\end{figure}

Figure~\ref{fig:IAA} shows \IAADphi, the ratio of \YtildepT(\dphi) distributions measured in central \AuAu\ and \pp\ collisions, for \gammadir\ and \pizero\ triggers and recoil jets with $\rr=0.2$ and 0.5. For \gammadir\ triggers the binning is different for central \AuAu\ and \pp\ collisions, due to different dataset sizes. The denominator of \IAADphi\ is therefore determined by fitting an exponential function to the \pp\ spectrum and interpolating. The smoothly--varying systematic uncertainty is likewise interpolated. For the \pp\ data points in Fig.~\ref{fig:DelphiDist} which show a limit, Figure~\ref{fig:IAA} utilizes the lower limit of the systematic uncertainty for the \AuAu\ data for the numerator in the ratio. The uncertainty boxes are the quadrature sum of uncorrelated uncertainties in numerator and denominator; these residual uncertainties in the ratio are nevertheless correlated between different \dphi\ bins. Fig.~\ref{fig:IAA}, right panels, show \IAADphi\ integrated over $3\pi/4<\dphi<\pi$, reported as \IAA\ in Ref.~\cite{STAR:2023pal}; these measurements are consistent. 

For large acoplanarity, all panels show suppression of \IAADphi\ for recoil jets with $\rr=0.2$ and significant enhancement for $\rr=0.5$. The value of \IAADphi\ at $\dphi\approx 2.65$ differs for $\rr=0.2$ and 0.5 by a factor $20\pm2$ (sys) for \pizero\ triggers and recoil jet $15<\pTjetch<20$ \gev\ (bottom panel), and a factor $3.8\pm1.7$ for \gammadir\ triggers and recoil jet $10<\pTjetch<15$ \gev\ (top panel); statistical error is negligible. Significant differences are likewise observed for $\dphi\approx 2.35$. A similar, marked \rr--dependent broadening of \IAADphi\ was observed in the same \pTjetch-range for h+jet correlations in central \PbPb\ collisions at $\sqrtsNN=5.02$ TeV~\cite{ALICE:2023qve}. 

Medium--induced yield enhancement at large acoplanarity may arise from secondary partonic scattering with QGP quasiparticles~\cite{Appel:1985dq,Blaizot:1986ma,DEramo:2010wup,DEramo:2012uzl,Chen:2016vem,DEramo:2018eoy}. However, such scattering effects should be evident for all \rr--values, which are used to probe the population of hard-scattering processes with different apertures. In contrast, Fig.~\ref{fig:IAA} and Ref.~\cite{ALICE:2023qve} show selective enhancement for $\rr=0.5$ but not $\rr=0.2$, which is not consistent with jet scattering as the predominant underlying mechanism. 
 
Another potential source of acoplanarity broadening is MPIs, as discussed above. While the MPI contribution to \IAApT\ is negligible at RHIC energies~\cite{STAR:2017hhs}, MPI may still affect the tail of the \IAADphi\ distribution. However, the strong \rr-dependence of  \IAADphi\ in Fig.~\ref{fig:IAA} likewise disfavors that scenario, since MPI effects should also broaden the acoplanarity distribution for all \rr\ values.

A consistent picture accommodating these observations is that selective acoplanarity broadening for large \rr\ and low \pTjetch\ arises from medium response, whereby the trigger--correlated ``jets'' observed at large angular deviation from $\dphi\approx \pi$ represent the diffuse wake or medium response to a recoil jet propagating in the QGP~\cite{ALICE:2023qve,ALICE:2023jye}. Figure~\ref{fig:IAA} therefore shows evidence of the medium response to the passage of an energetic jet, which has not been identified previously at RHIC.

The \gammadir--triggered and \pizero--triggered distributions are expected to differ in recoil jet relative  quark/gluon fraction and in average in-medium path length~\cite{STAR:2023ksv,STAR:2023pal}. The medium--induced azimuthal broadening shown in Fig.~\ref{fig:IAA} is qualitatively similar for \gammadir\ and \pizero\ triggers, though the distributions differ in detail. Systematic comparison of these two distributions with model calculations may provide new insight into the color charge and pathlength dependence of jet quenching effects.

Figure~\ref{fig:IAA} shows comparisons of several theoretical calculations incorporating jet quenching with the data: JEWEL~\cite{Zapp:2013vla,Zapp:2008gi}, with medium--recoil effects; the analytic QCD calculation shown in Fig.~\ref{fig:DelphiDist}, with Gaussian distributed in--medium broadening~\cite{Chen:2016vem}; and the Hybrid Monte Carlo model~\cite{Casalderrey-Solana:2014bpa,Casalderrey-Solana:2016jvj,Casalderrey-Solana:2018wrw} with hydrodynamic wake implemented. These calculations are not smeared by the \ETtrig\ resolution, whose effects are similar in \pp\ and \AuAu\ and largely cancel in the \IAADphi\ ratio~\cite{STAR:2023ksv}.

JEWEL is based on PYTHIA, which describes the \pp\ measurements well (Fig.~\ref{fig:DelphiDist}). JEWEL also describes well the \rr--dependent medium--induced acoplanarity broadening seen at the LHC~\cite{ALICE:2023qve}. However, at RHIC energies JEWEL does not exhibit significant medium--induced broadening for either $\rr=0.2$ and 0.5, for both \gammadir\ and \pizero\ triggers (Fig.~\ref{fig:IAA}).

For the analytic QCD calculation the Gaussian broadening width is \qhatL, where \qhat\ is the jet transport coefficient~\cite{Majumder:2010qh}, $L$ is the in--medium path length, and $\langle{\ldots}\rangle$ indicates averaging over collisions. The band in Figure~\ref{fig:IAA} corresponds to $3<\qhatL<13$ GeV$^{2}$, reproducing the measured \IAADphi\ for $\rr=0.5$, but not for $\rr=0.2$. This feature is intrinsic to the simple Gaussian azimuthal broadening employed, which models jet--QGP multiple scattering without \rr\ dependence. This disagreement with data provides additional evidence that the observed marked \rr--dependence does not arise predominantly from in--medium soft scattering.

The Hybrid Model predicts medium--induced narrowing for both $\rr=0.2$ and $\rr=0.5$ (Fig.~\ref{fig:IAA}), in disagreement with the data, with similar predictions at LHC energies~\cite{ALICE:2023qve}. The End Matter presents Hybrid Model calculations for 
recoil jet $5<\pTjetch<10$ \gev\ which exhibit \rr--dependent broadening whose qualitative features are similar to those seen in data for larger values of \pTjetch, but only with wake implemented. This provides additional insight into the physical origin of the broadening, and its modeling. 

\section{Summary} This article reports measurements of the semi--inclusive acoplanarity distribution of charged-particle jets recoiling from \gammadir\ and \pizero\ triggers in \pp\ and central \AuAu\ collisions at $\sqrtsNN=200$ GeV. Significant \rr--dependent medium--induced acoplanarity broadening is observed, corresponding to a yield enhancement for large \rr\ compared to small \rr\ jets of up to a factor 20. A picture that accommodated these observations and a corresponding measurement at the LHC is that the broadening arises predominantly from diffuse QGP medium response to the passage of an energetic parton, i.e. the jet wake, rather than single hard  Rutherford--like scattering off of QGP quasiparticles. Theoretical calculations incorporating jet quenching and QGP wake effects exhibit significant differences with the measurements, requiring modification of their underlying physics description to improve the agreement. These measurements provide new insight into the nature of the interaction between jets and the QGP, and the application of jets to probe QGP dynamics. 

\begin{acknowledgments}
 We thank Jaime Norman, Yu Shi, Shu-Yi Wei, Bowen Xiao, Feng Yuan, Danny Pablos, and Krishna Rajagopal for providing calculations. We thank the RHIC Operations Group and SDCC at BNL, the NERSC Center at LBNL, and the Open Science Grid consortium for providing resources and support.  This work was supported in part by the Office of Nuclear Physics within the U.S. DOE Office of Science, the U.S. National Science Foundation, National Natural Science Foundation of China, Chinese Academy of Science, the Ministry of Science and Technology of China and the Chinese Ministry of Education, NSTC Taipei, the National Research Foundation of Korea, Czech Science Foundation and Ministry of Education, Youth and Sports of the Czech Republic, Hungarian National Research, Development and Innovation Office, New National Excellency Programme of the Hungarian Ministry of Human Capacities, Department of Atomic Energy and Department of Science and Technology of the Government of India, the National Science Centre and WUT ID-UB of Poland, German Bundesministerium f\"ur Bildung, Wissenschaft, Forschung and Technologie (BMBF), Helmholtz Association, Ministry of Education, Culture, Sports, Science, and Technology (MEXT), and Japan Society for the Promotion of Science (JSPS).
\end{acknowledgments}

\clearpage
\appendix

\setcounter{secnumdepth}{2}
\renewcommand{\thefigure}{\Alph{section}\arabic{figure}}
\setcounter{figure}{0}

\section{Hybrid Model comparison} 
\label{sect:HybridModelAppendix}
 \begin{figure}[htb!]
 \centering
\includegraphics[width=0.5\textwidth]{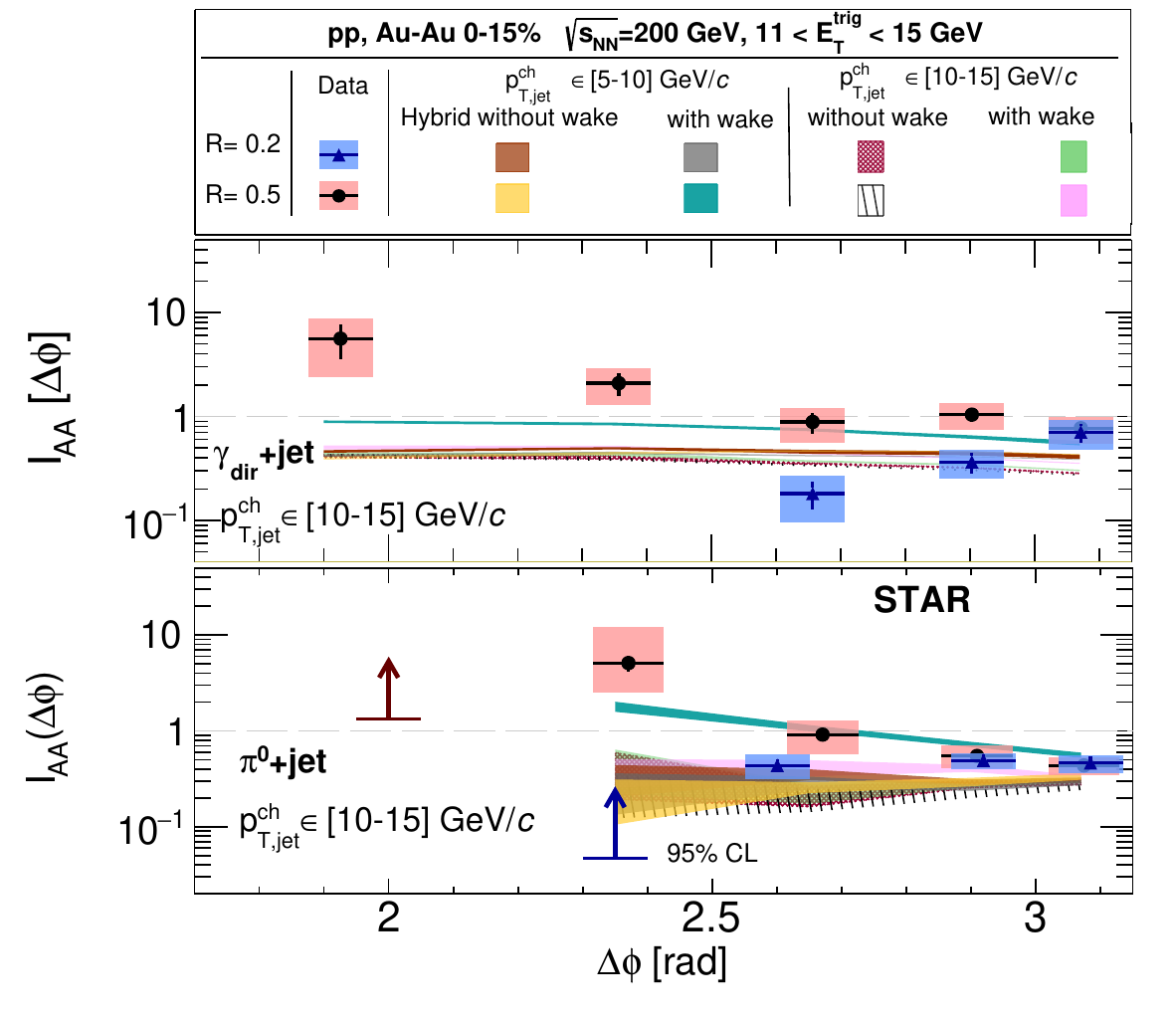}
\caption{Data from Fig.~\ref{fig:IAA} for recoil jet $10<\pTjetch<15$ \gev, and for Hybrid Model calculations for both $10<\pTjetch<15$ \gev\ and $5<\pTjetch<10$ \gev, both with and without wake.}
\label{fig:HybridComp}
 \end{figure}

Figure~\ref{fig:HybridComp} shows the \gammadir\ and \pizero--triggered \IAADphi\  distributions for recoil jet $10<\pTjetch<15$ \gev\ (Fig.~\ref{fig:IAA}), and for Hybrid Model calculations for $10<\pTjetch<15$ \gev\ and $5<\pTjetch<10$ \gev, both with and without wake. For both triggers, Hybrid Model calculations exhibit no significant \rr-dependence without wake in both \pTjetch\ intervals, or with wake for $10<\pTjetch<15$ \gev. However, for $5<\pTjetch<10$ \gev, the calculations with wake exhibit a marked \rr-dependent acoplanarity broadening for both triggers.  Although this is qualitatively similar to the effect seen in data for $10<\pTjetch<15$ \gev, the theoretical model shows a larger effect for \pizero\ triggers than \gammadir\ triggers. 

The striking \rr-dependent medium--induced acoplanarity broadening seen in data is therefore reproduced qualitatively by the Hybrid Model, though only with the hydrodynamic wake implemented and only in a lower kinematic interval than the reported measurement.

\section{Variation of ME normalization region}
\label{sect:AppedixMEnorm}

\begin{figure}[htb!]
 \centering
\includegraphics[width=0.5\textwidth]{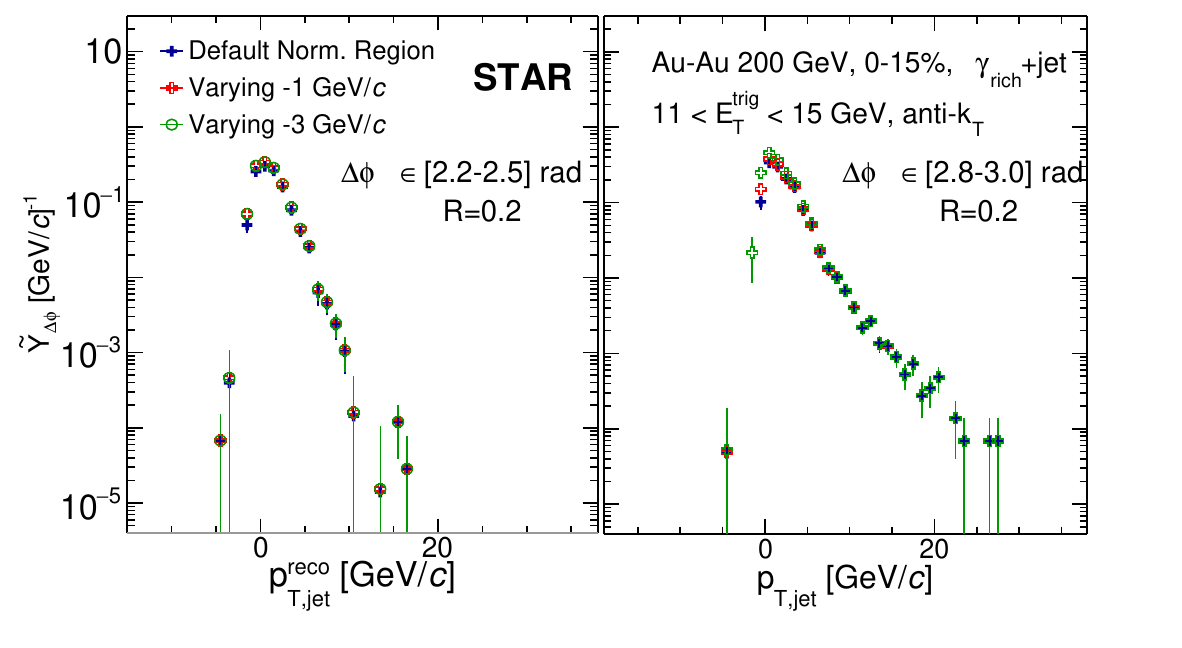}
\includegraphics[width=0.5\textwidth]{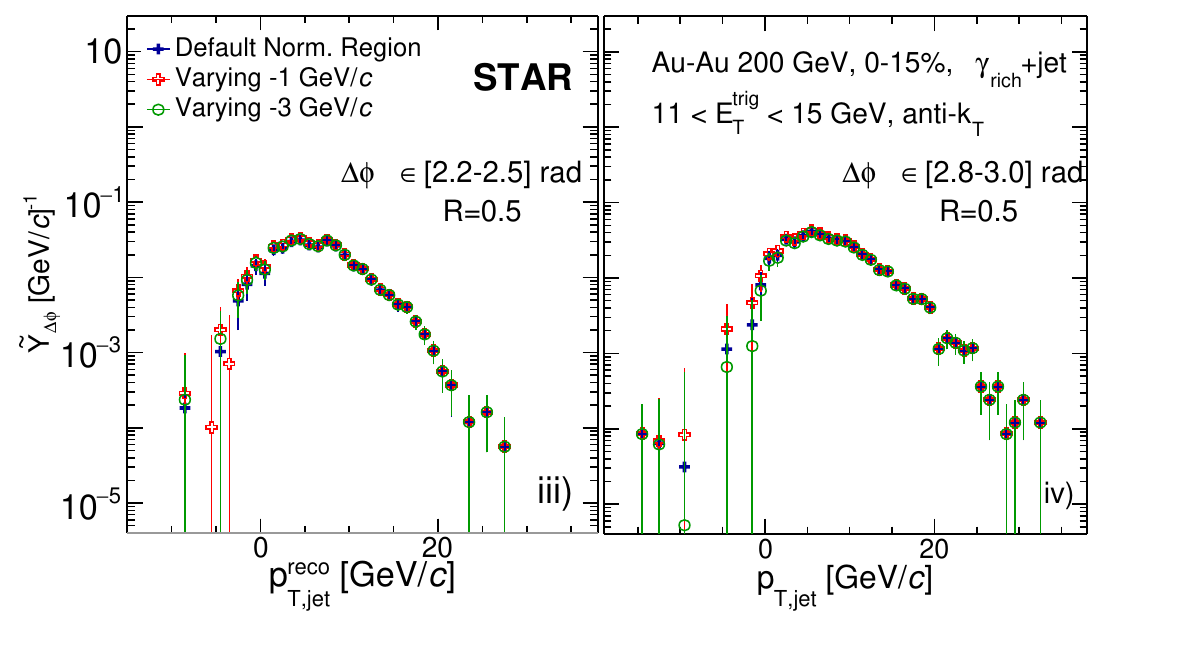}
\caption{Distributions of \Ytildedphi(\pTreco) from PRL Fig. [3] using the nominal ME normalization region (blue), and those where the upper limit of the normalization region are reduced by 1 (red) or 3 (green) \gev.}
\label{fig:YtildepTVariation}
\end{figure}

This section documents the systematic uncertainty due to variation in the limits of the ME normalization region. Figure~\ref{fig:YtildepTVariation} shows the distribution of \Ytildedphi(\pTreco) (Eq.~[3]) for the nominal ME normalization region (Fig.~[1]) and for shifting the upper limit of the normalization region lower by 1 and 3 \gev. The resulting variation in \Ytildedphi(\pTreco) is seen to be small, particularly for $\pTreco>0$, but is nevertheless incorporated into the systematic uncertainty of the measurement.

\section{Effect of \dphi-smearing correction}

\begin{figure}[htb!]
 \centering
\includegraphics[width=0.5\textwidth]{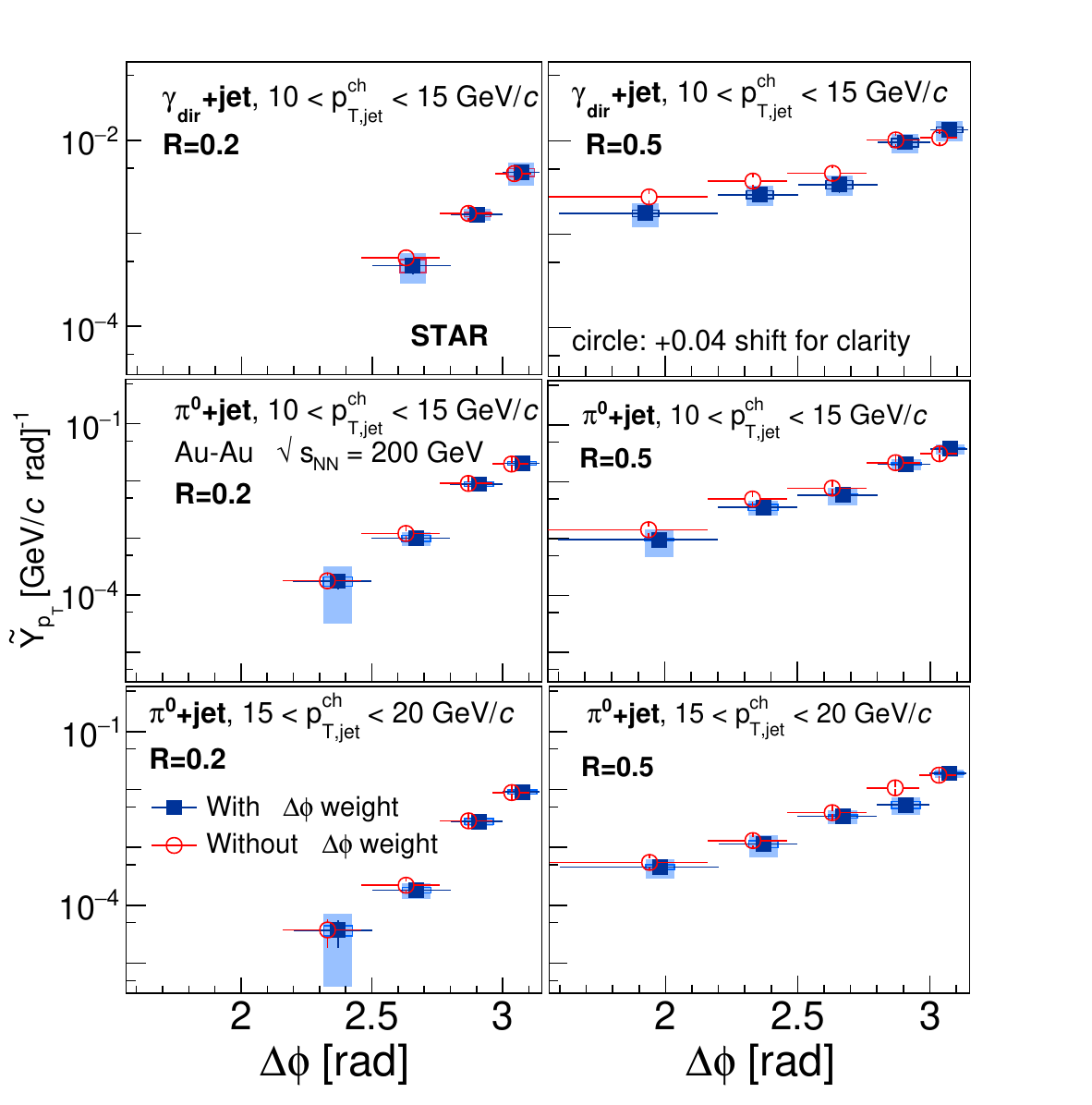}
\caption{Distributions of \YtildepT(\dphi) with bin-wise correction for \dphi-smearing from Fig.~[2] (filled boxes), and the central value of the distributions without the bin-wise correction (open circles).}
\label{fig:dphiCorrection}
\end{figure}

This section documents the bin-wise correction for \dphi-smearing applied to the measurement of  \YtildepT(\dphi) . Figure~\ref{fig:dphiCorrection} shows \YtildepT(\dphi) distributions with the bin-wise correction (Fig.~[2])), and the central value of the same distributions without the bin-wise correction.

\begin{figure}[htb!]
 \centering
\includegraphics[width=0.5\textwidth]{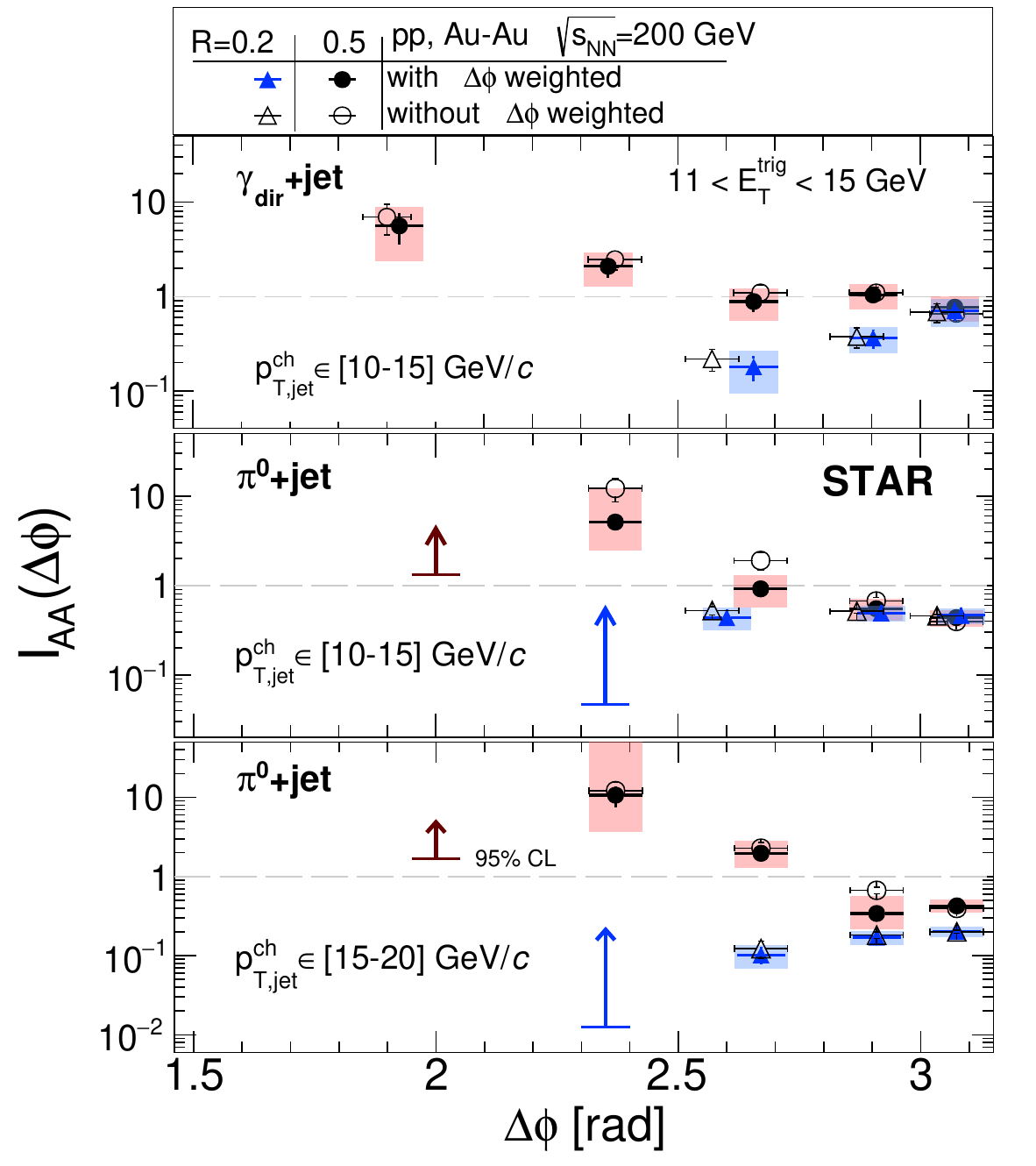}
\caption{Distributions of \IAADphi\ (Fig.~[3]) (colored markers) compared to the central values of the same ratios without the bin-wise correction for \dphi-smearing.}
\label{fig:dphiCorrectionRatio}
\end{figure}

Figure \ref{fig:dphiCorrectionRatio} shows \IAADphi\ with the bin-wise correction (Fig.~[3]), and its central value without the bin-wise correction. Figures~\ref{fig:dphiCorrection} and \ref{fig:dphiCorrectionRatio} show that the magnitude of the bin-wise correction for smearing in \dphi\ is sub-leading relative to other effects contributing to the systematic uncertainty, and is negligible in comparison to the magnitude of medium-induced modification observed in the measurements of \IAADphi.



\bibliographystyle{apsrev4-2}


\bibliography{references}

\end{document}